\documentstyle[preprint,aps] {revtex}

\begin{document}

\title {\bf Why hyperbolic theories of dissipation cannot be ignored: Comments
on a paper by Kost\"{a}dt and Liu}

\author{Luis Herrera\thanks{ Also at UCV, Caracas, Venezuela; E-mail address:
lherrera@gugu.usal.es}\\
Area de F\'{\i}sica Te\'orica, Facultad de Ciencias,\\ Universidad de
Salamanca, 37008 Salamanca, Spain.\\
and\\
Diego Pav\'{o}n\thanks{E-mail address: diego@ulises.uab.es}\\
Departamento de F\'{\i}sica, Facultad de Ciencias,\\
Edificio Cc, Universidad Aut\'{o}noma de Barcelona,\\
08193 Bellaterra, Spain.
}

\date{}
\maketitle

\begin{abstract}
Contrary to what is asserted in a recent paper by Kost\"{a}dt and Liu
(``Causality and stability of the relativistic diffusion equation") \cite{Liu},
experiments can tell apart (and in fact do) hyperbolic theories from
parabolic theories of dissipation. It is stressed that
the existence of a non--negligible relaxation time does not imply for
the system to be out of the hydrodynamic regime.
\end{abstract}

\newpage

As is well--known hyperbolic theories of fluid dissipation were formulated
to get rid of some undesirable features of parabolic theories, such
as acausality \cite{Israel}. This was achieved at the price of extending the
set  of field variables by including the dissipative fluxes (heat current,
nonequilibrium stresses and so on) at the same footing as the old ones
(energy densities, equilibrium pressures, etc), thereby giving rise to
more physically satisfactory but involved theories from the mathematical
point of view. A key quantity in these theories is the relaxation time $\tau$
of the corresponding  dissipative process. This positive--definite quantity
has a distinct physical meaning, namely the time taken by the system to return
spontaneously to the steady state (whether of thermodynamic equilibrium or
not) after it has been suddenly removed from it. It is, however,  somehow
connected to the mean collision time $t_{c}$ of the  particles responsible
for the dissipative process, ofentimes erroneously identified with it.
In principle they are different since $\tau$ is (conceptually
and many times in practice) a macroscopic time, although in some instances
it may correspond just to a few $t_{c}$. No general formula linking
$\tau$ and $t_{c}$ exists, the relationship between them depends in
each case on the system under consideration. As mentioned above,
it is therefore appropriate to interpret $\tau$ as the time taken by
the corresponding dissipative flow to relax to its steady value.

Thus, it is well known that the classical Fourier law for heat current,
leads to a parabolic equation for temperature (diffusion equation), which
does not forecast propagation of perturbations along characteristic causal
cones (see \cite{Joseph}, \cite{Jou}, \cite{Maartens} and references
therein). In other words perturbations propagate with infinite speed.
This non--causal behavior is easily visualized, by taking a look on the
thermal conduction in an infinite medium (see \cite{Landau}). The
origin of this behavior is to be found in the parabolic character of
Fourier's law, which implies that the heat flow starts (vanishes)
simultaneously with the appearance (dissapearance) of a temperature
gradient. Although $\tau$ is very small for phonon-electron, and
phonon-phonon interaction at room temperature (${\cal O}(10^{-11})$ and
${\cal O}(10^{-13})$ sec, respectively \cite{Peierls}), neglecting it
is the source of difficulties, and in some cases a bad approximation
as for example in superfluid  Helium \cite{Peshkov}, and degenerate
stars where thermal conduction is dominated by electrons -see
\cite{Joseph}, \cite{Jou}, \cite{mdv}, for further examples.

In order to overcome this problem many researchers, starting with Cattaneo
and Vernotte \cite{Cattaneo}, generalized the Fourier law by introducing
a relaxation time, thereby leading to a hyperbolic equation for the
temperature.

Obviously, $\tau$ shouldn't be neglected if one wishes to study transient
regimes, i.e., the departure from a initial steady situation and the
approach to the a new one.
In fact, leaving aside the problem of stability and the fact that parabolic
theories are necessarily non--causal, it is obvious that whenever
the time scale of the problem under consideration is of the order of
(or smaller) than the relaxation time, the latter cannot be ignored.
It is common sense what is at stake here: neglecting the relaxation time
ammounts -in this situation- to disregard the whole problem under
consideration. Such a neglecting literally means to throw the baby
with the  water!

In a recent paper by Kost\"{a}dt and Liu \cite{Liu}, arguments have been put
forward suggesting that parabolic theories of dissipation are healthy
enough, and that hyperbolic (i.e., causal) theories are not necessary when
dealing with dissipative fluid systems. In particular these authors state that

\begin{quote}
In fact, recently, it has been shown by
Geroch \cite{Geroch} and Lindblom \cite{Lindblom} that the complicated
dynamical structure which ensures causality is unobservable. The evolution of
any physical fluid state according to any causal theory results in
energy--momentum tensors and particle currents that are experimentally
indistinguishable from the respective hydrodynamic expressions.
\end{quote}

We would like to stress that the quoted phrase is at variance with
experimental evidence as a number observations unambiguosly show
\cite{mdv}. The aim of this Comment is to indicate the roots of the
confusion leading to that erroneous view \cite{save}.


The basic assumption underlying the disposal of hyperbolic
dissipative theories, states that systems with relaxation
times comparable to the characteristic time of the system
are out of the hydrodynamic regime \cite{regime}.
This can be valid only if the particles making up the fluid are
the same ones that transport the heat. However, this is (almost?)
never the case. Specifically, for a neutron star, $\tau$ is of the
order of the scattering time between electrons (which carry the
heat) but this fact is not an obstacle (no matter how large the
mean free path of these electrons may be) to consider the neutron
star as formed by a Fermi fluid of degenerate neutrons. The same
is true for the second sound in superfluid Helium and solids, and
for almost any ordinary fluid. In brief, the hydrodynamic regime
refers to fluid particles that not necessarily (and as a matter of fact,
almost never) transport the heat. Therefore large relaxation times (large
mean free paths of particles involved in heat transport) does not imply a
departure from the hydrodynamic regime (this fact has been streseed before
\cite{Santos}, but is usually overlooked).

However, even in the case when particles that make up the fluid are
responsible of the dissipative process, the taking for granted that
$\tau$ and $t_{c}$ are {\it always} of the same order, or
what comes to the same that the dimensionless quantity
$\Gamma \equiv (\tau c_{s}/L)^{2}$ is negligible in all instances
\cite{Geroch}, \cite{Lindblom}, is not always valid -here $c_{s}$
stands for the adiabatic speed of sound in the fluid under
consideration and $L$ the characteristic length of  the
system. That assumption would be right if $\tau$ were always
comparable to $t_{c}$ and $L$ always ``large",
but there are, however, important situations in which
$\tau \gg t_{c}$, and  $L$ ``small" although still large enough to
justify a macroscopic description. For tiny semiconductor pieces of
about $10^{-4}$ cm in size, used in common electronic
devices submitted to high electric fields, the
above dimensionless combination (with $\tau \sim 10^{-10}$
sec, $c_{s} \sim 10^{7}$ cm/sec \cite{muscato}) can easily
be of the order of unity. In ultrasound propagation as well as
light-scattering experiments in gases and  neutron-scattering in
liquids the relevant length is no longer the system size,
but the wavelenght $\lambda$ which is usually much
smaller than $L$ \cite {Weiss}, \cite{Copley}.
Because of this, hyperbolic theories may bear some
importance in the study of nanoparticles and quantum dots.
Likewise in polymeric fluids relaxation
times are related to the internal configurational
degres of freedom and so much longer than $t_{c}$
(in fact they are in the range of the minutes), and
$c_{s} \sim 10^{5}$ cm/sec, thereby $\Gamma \sim {\cal O}(1)$.
In the degenerate core of aged stars the
thermal relaxation time can be as high as $1$ second
\cite{Harwit}. Assuming the radius of the core of about
$10^{-2}$ times the solar radius, one has $\Gamma \sim {\cal O}(1)$
again. Fully ionized plasmas exhibit a collisionless regime
(Vlasov regime) for which the parabolic hydrodynamics predicts
a plasmon dispersion relation at variance with the microscopic results;
the latter agree, however, with the hyperbolic hydrodynamic approach
\cite{Tokatly}. Think for instance of some syrup fluid flowing
under a imposed shear stress, and imagine that the shear is suddenly
switched off. This liquid will come to rest only after a much longer
time ($\tau$) than the collision time between its constituent
particles has elapsed. Many other examples could be added
but we do not mean to be exhaustive.

Even in the steady regime the descriptions offered by causal
and acausal theories do not necessarily coincide.
The differences between them in such a situation
arise from (i) the presence of $\tau$ in terms that
couple the vorticity to the heat flux and shear stresses.
These may be large even in steady states (e.g. rotating
stars). There are also other acceleration coupling terms
to bulk and shear stresses and heat flux. The coefficients
for these vanish in parabolic theories, and they could
be large even in the steady state. (ii) From the convective
part of the time derivative (which are not negligible
in the presence of large spatial gradients). (iii) From
modifications in the equations of state due to the presence of
dissipative fluxes \cite{Jou}.

However, it is precisely before the establishment of the steady
regime that both types of theories (hyperbolic and parabolic)
differ more importantly. It is well--known (see  \cite{Joseph},
\cite{Jou}, \cite{mdv}, \cite{Herrera}) that a variety of physical
processes take place on time scales of the order of (or even smaller)
than the corresponding relaxation time, which as was stressed above
does not imply that the system is out of hydrodynamic regime.
Therefore if one wishes to study a dissipative process for times
shorter than $\tau$, it is mandatory to resort to a hyperbolic
theory which is a more accurate macroscopic approximation
to the underlying kinetic description.

Only for times longer than $\tau$ it is permissible
to go to a parabolic one, provided that the spatial
gradients are not so large that the convective part of the time
derivative does not become important, and that the fluxes and
coupling terms remain safely small. But even in these cases, it should
be kept in mind that the way a system leaves the equilibrium may
critically depends upon relaxation time \cite{Herrera}.
Therefore the future of the system at time scales much longer than the
relaxation time (once the steady state is reached),
may also critically depend on $\tau$.

Thus, even though parabolic theories have proved very useful for many
practical purposes, it appears that there are a number of well-known
instances (such as transient regimes) where they fail hopelessly, but
hyperbolic theories sucessfully predict the experimental results -i.e.,
they are distinguishable. Having said this, it is worth mentioning
that at the moment it is rather uncertain which among the proposed
hyperbolic theories [2] will eventually emerge as ``the correct one".
This discrimination seems to lay a long way ahead.

We hope this Comment will help to convince the reader that hyperbolic
theories are indeed of not mere academic interest and it wouldn't be
wise to dispense of them.

\section*{Acknowledgements}
The authors are indebted to David Jou for reading the manuscript and
helpful remarks. This work has been partially supported by the Spanish
Ministry of Education under grants PB94-0718 and PB96-1306.


\begin{thebibliography}{88}
\bibitem{save} It is not our purpose to challenge the proof
offered by Kost\"{a}dt and Liu on the stability of parabolic
equations, only to focus on the unfortunate quoted phrase
which, on the other hand, is irrelevant to the stability
proof.
\bibitem{regime}
The concept of hydrodynamic regime invokes the ratio between the mean
free path of fluid particles  and the
characteristic length of the system. When this ratio is lower that unity,
the fluid is within the hydrodynamic regime. When it is larger than unity,
the regime becomes Knudsen's. In the latter case the fluid ceases to be
a continuum and even hyperbolic theories are no longer realiable.
\end{thebibliography}
\end{document}